\documentclass[runningheads]{llncs}
\usepackage{graphicx}
\usepackage{amsfonts}
\usepackage{amsmath}
\usepackage{makecell}

\usepackage{threeparttable}
\usepackage{multirow}
\usepackage{booktabs}
\usepackage{bbm}
\usepackage{bbold}
\usepackage{todonotes}
\usepackage{subcaption}
\usepackage{bbding}

\begin{document}
\title{Max-Fusion U-Net for Multi-Modal Pathology Segmentation with Attention\\and Dynamic Resampling}

\author{Haochuan Jiang\inst{1}
\and Chengjia Wang\inst{2}\Envelope
\and Agisilaos Chartsias \inst{1}
\and Sotirios A. Tsaftaris\inst{1,3}}

\authorrunning{H. Jiang et al.}
\titlerunning{Max-Fusion U-Net for Multi-Modal Pathology Segmentation}

\institute{School of Engineering, University of Edinburgh,  U.K.
\and Center for Cardiovascular Science, University of Edinburgh, U.K., \email{chengjia.wang@ed.ac.uk}
\and The Alan Turing Institute, U.K.}

\maketitle              

\begin{abstract}
Automatic segmentation of multi-sequence (multi-modal) cardiac MR (CMR) images plays a signiﬁcant role in diagnosis and management for a variety of cardiac diseases. However, the performance of relevant algorithms is signiﬁcantly aﬀected by the proper fusion of the multi-modal information. Furthermore, particular diseases, such as myocardial infarction, display irregular shapes on images and occupy small regions at random locations. These facts make pathology segmentation of multi-modal CMR images a challenging task. In this paper, we present the Max-Fusion U-Net that achieves improved pathology segmentation performance given aligned multi-modal images of LGE, T2-weighted, and bSSFP modalities. Speciﬁcally, modality-speciﬁc features are extracted by dedicated encoders. Then they are fused with the pixel-wise maximum operator. Together with the corresponding encoding features, these representations are propagated to decoding layers with U-Net skip-connections. Furthermore, a spatial-attention module is applied in the last decoding layer to encourage the network to focus on those small semantically meaningful pathological regions that trigger relatively high responses by the network neurons. We also use a simple image patch extraction strategy to dynamically resample training examples with varying spacial and batch sizes. With limited GPU memory, this strategy reduces the imbalance of classes and forces the model to focus on regions around the interested pathology. It further improves segmentation accuracy and reduces the mis-classiﬁcation of pathology. We evaluate our methods using the Myocardial pathology segmentation (MyoPS) combining the multi-sequence CMR dataset which involves three modalities. Extensive experiments demonstrate the eﬀectiveness of the proposed model which outperforms the related baselines.
The code is available at https://github.com/falconjhc/MFU-Net.

\end{abstract}

\section{Introduction}\label{Sec:Introduction}
Cardiac diseases are typically assessed using multiple cardiac MR (CMR) sequences (modalities), providing complementary information. For example, Late Gadolinium Enhancement (LGE) detects myocardial infarct, T2-weighted (T2) images provide clear visibility of acute injury and ischemic regions, and balanced-Steady State Free Precession cine sequence (bSSFP) offers high contrast between anatomical regions and captures cardiac motion.

Deep learning models have been extensively used for automatic segmentation of multi-modal data. 
A critical step for the analysis of multi-modal CMR data is to effectively fuse information from multiple modalities. 
Prior works~\cite{havaei2017brain} concatenate the feature maps extracted from different modalities into different channels and fuse them in the following convolutional layers.
Other methods~\cite{mahmood2018multimodal,dolz2018hyperdense} merge the features across different layers of the neural network, where a cross-modal convolution fusion model is introduced in~\cite{tseng2017joint}.
In \cite{jiang2018w} they employ dedicated encoders for different modalities to encode different types of information, for example, content and style features from the corresponding input data. The features are then fused using channel concatenation in the U-Net skipping-connections. 
A similar idea was also used in~\cite{chartsias2019disentangle} where a maximum fusion operator instead of simple concatenation in the skip-connections is applied on disentangled anatomy factors extracted from different modalities at the end of encoders.

\begin{figure}[ht]
\begin{subfigure}{0.5\linewidth}
  \centering
  \includegraphics[width=\linewidth]{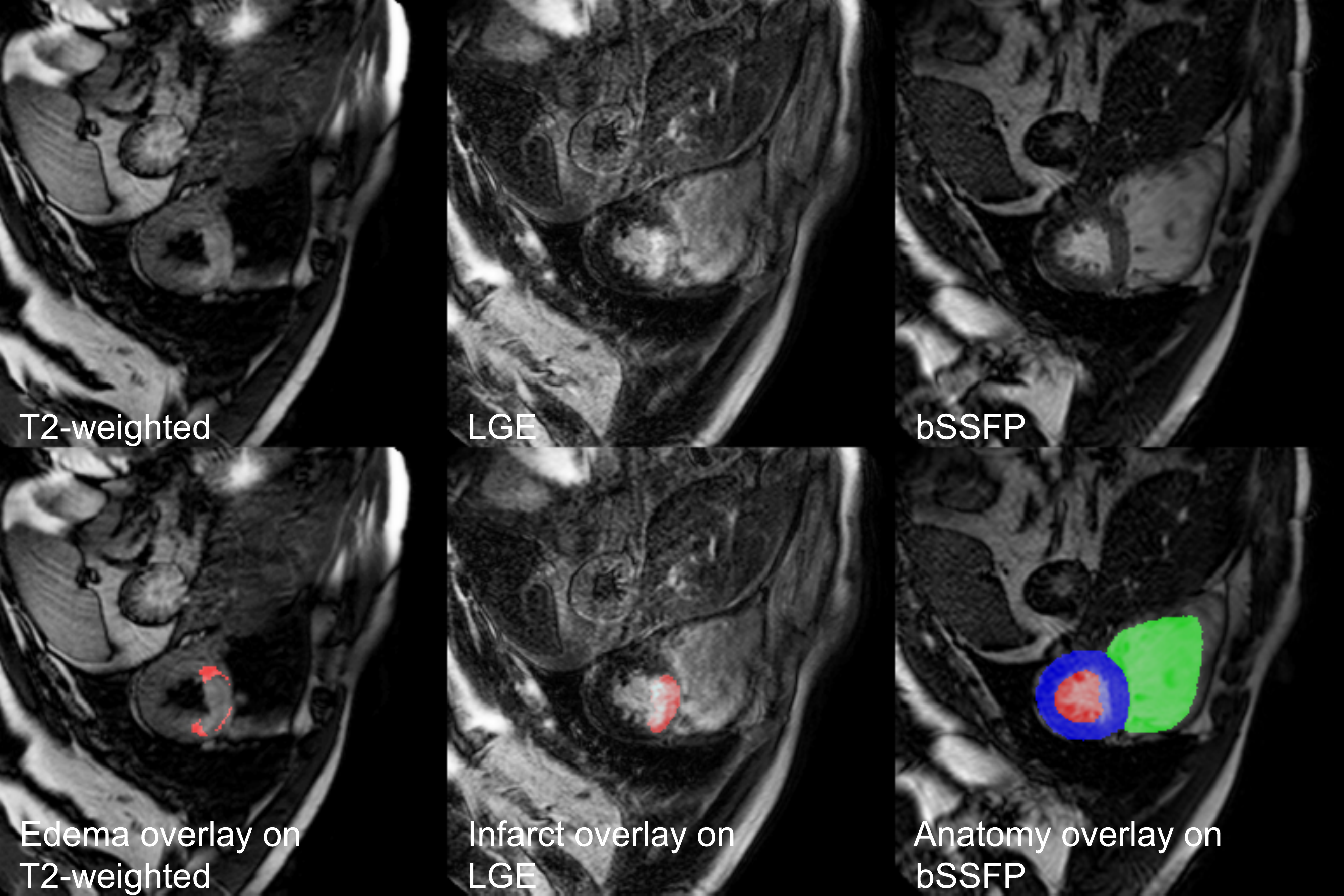}
  \caption{Example 1 with anatomy overlay}
  \label{fig:challenge1}
\end{subfigure}
\begin{subfigure}{0.5\linewidth}
  \centering
  \includegraphics[width=\linewidth]{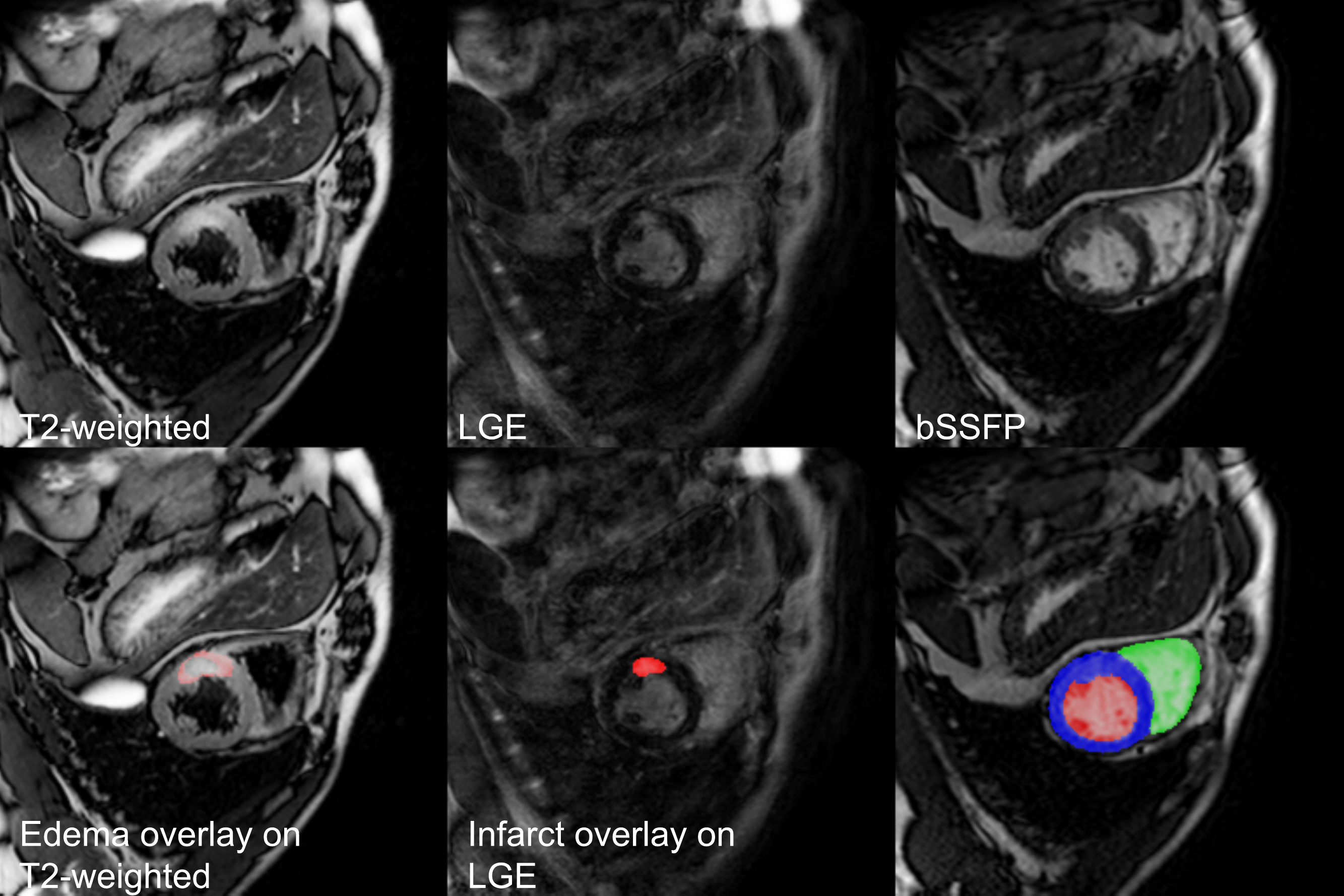}
  \caption{Example 2 with anatomy overlay}
  \label{fig:challenge2}
\end{subfigure}
\caption{Examples of multi-modal CMR images overlaying anatomy and pathology.}
\label{fig:challenge}
\end{figure}

One other challenge in segmenting pathology such as myocardial infarct and edema is that these pathologies are often of diverse shape and occur at random positions. 
As such, shape priors such as mask discriminator~\cite{chartsias2019disentangle} cannot be used. 
Besides, the interested pathology and anatomy only occur within a small region of the whole image, as examples of multi-sequence CMR images with manually segmented anatomy and pathology (myocardial infarction and edema) given in Fig.~\ref{fig:challenge}.
This makes the data distribution highly imbalanced across classes, resulting in overfitting in the training data.
Particularly in current popular backbone convolutional neural networks (CNN) assuming all pixels in the image contribute equally to the final prediction, the over-fitting issue is even worse.
A possible solution is to use the spatial attention module~\cite{fu2019dual}, leading the network to focus on specific image regions.
In our case, the focus corresponds to pathology pixels.

In addition, given limitations in GPU memory, training can only be performed with a small batch size. 
This even worsen the overfitting issue since due to this and small pathological region in each image, in each training iteration,  only a small amount of pathology pixels are seen by the network.
Nevertheless, the batch size can be increased if training with smaller size patches instead of full images, e.g., by engaging random cropping.
Although it is commonly used as a data augmentation technique~\cite{takahashi2018ricap}, all patches are treated equally importantly.
It is appealing if the cropping strategy will oversample patches around pathology regions that we are interested in.

In this paper, we propose the Max-Fusion U-Net (MFU-Net) for cardiac pathology segmentation, given fully-annotated multi-modal aligned images.
We use dedicated encoders to extract features for each modality, as in~\cite{chartsias2019disentangle,jiang2018w}.
But rather than channel concatenation~\cite{jiang2018w}, we fuse features from different modalities with the pixel-wise maximum operator applied on each layer~\cite{chartsias2019disentangle}. 
This fusion operator guides the network to keep informative features extracted by each modality.
At the same time, fusion with maximum operator indirectly encourages feature maps to encode important features in high intensities including pathological pixels.
A spatial-attention module is also employed in the last decoding layer to modulate the spatial focus, which in our case means to increase focus of the pathology pixels. 
Finally, to address the issue that only a small amount of pathological pixels are exposed to the network during training, we adopt a dynamic resampling strategy. To obtain each batch, we extract multiple patches around the interested pathology based on an arbitrary probability, then extract the rest data by randomly cropping the image to the same size. 
By feeding more patches related to pathological regions and less related to background patches, the network will thus naturally become more sensitive to pathological pixels. 
At the same time, the training batch size can be dynamically enlarged without occupying extra computation resources due to the reduced image dimension. 
Theoretically, the spatial size of the training data should not harm the training efficiency as long as the sampled image patches are bigger than the largest receptive field of the network. 
Extensive experiments have demonstrated the effectiveness of the proposed MFU-Net in cardiac pathology segmentation including infarction and edema when given multi-modal inputs including LGE, T2-weighted, and bSSFP, outperforming relevant methods. 
Major \textbf{contributions} of this work are summarized as follows:
\begin{itemize}
	\item We proposed the MFU-Net that fuses multi-modal features extracted by dedicated encoders with the pixel-wise maximum operator;
	\item We incorporate a spatial-attention module to guide the network to focus on the pathology region;
	\item We proposed a novel training strategy by feeding randomly resampled sub-patches from the original training data with more probability around the pathology region, at the same time increasing the batch size dynamically;
	\item MFU-Net improves the Dice score of state-of-the-art benchmarks on myocardial pathology segmentation on multi-modal CMR 2020 dataset~\cite{zhuang2016multivariate,zhuang2018multivariate}.
\end{itemize}

\section{Methodology}\label{Sec:Methodology}
This section presents the proposed MFU-Net model, and the details about the architecture, the modality-specific encoders, the maximum fusion operator, and the attention-based decoding modules.

\noindent\textbf{Overview:}
Let $X_{LGE}, X_{T2}, X_{bSSFP}$ represent images of LGE, T2-weighted, and bSSFP CMR modalities respectively, and $Y_{ana}$, $Y_{pat}$ be the associated anatomy and pathology masks.
If $i$ enumerates all samples from the above sets, we assume a fully labelled multi-modal pathology subset $ \mathbf{L} = \{x^i_{LGE}, x^i_{T2}, x^i_{bSSFP},y^i_{ana}, y^i_{pat}\}$, where three modality slices $x^i_{LGE}, x^i_{T2}, x^i_{bSSFP} \subset \mathbb{R}^{H\times W}$ are preprocessed~\cite{zhuang2016multivariate,zhuang2018multivariate}, such that they are aligned  in a common space and are resampled to the same spatial resolution.
In addition, $y_{ana}^i \in Y_{ana} := \{0,1\}^{H \times W \times N}$, and $y^i_{pat} \in Y_{pat} := \{0, 1\}^{H \times W \times K}$, where $N$ and $K$ denote the number of anatomy, and pathology masks respectively.\footnote {We restrict to the case where $N=3$ (myocardium, left ventricle, and right ventricle) and $K=2$ (infarction and edema).}, and $H$ and $W$ are the image height and width.

\begin{figure}[t!]
    \centering
    \includegraphics[width=\linewidth]{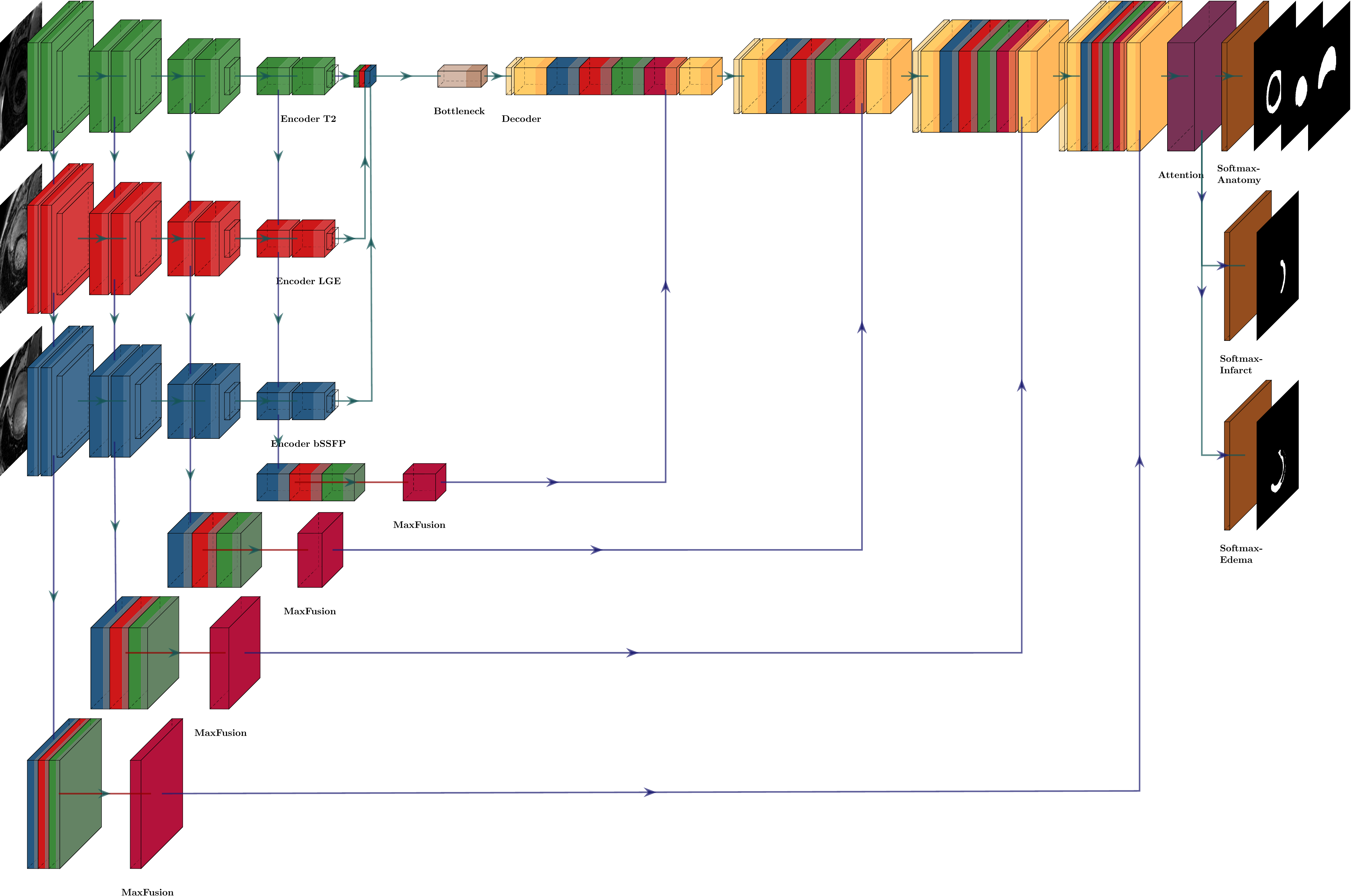}
    \caption{MFU-Net Architecture. Red, Green, and Blue blocks represent LGE, bSSFP, and T2 encoding features. Yellow blocks depict the decoding features. Pink Blocks are max-fused features, while transparent brown block is the bottleneck feature. Solid brown ones are the softmaxed probability map, while the amaranth block is the spatial attention module.}\label{fig:architecture}
\end{figure}

\subsection{Model Architecture} \label{sec:model_architecture}
The architecture of MFU-Net is illustrated in Fig.~\ref{fig:architecture}. It consists of three modality-specific encoders, a multi-modal feature fusion with pixel-wise maximum operator, and a decoder with a spatial attention module that produces the segmentation results.

\noindent\textbf{Individual Encoders:}
The original U-Net architecture~\cite{ronneberger2015u} only specifies a single encoder to extract features.
To accommodate differences in the pixel intensity distributions between modalities, we expand the U-Net by using one independent encoder for each modality. This leads to three modality-specific encoders.
Represented by red, green, and blue colors in Fig.~\ref{fig:architecture}, these encoders are denoted as $Enc_{LGE}$, $Enc_{T2}$, and $Enc_{bSSFP}$ respectively for LGE, T2, and bSSFP data.
The encoded features $Enc_{LGE}(x^i_{LGE})$, $Enc_{T2}(x^i_{T2})$, and $Enc_{bSSFP}(x^i_{bSSFP})$ are concatenated and used as input to the bottleneck blocks (the transparent brown blocks in Fig.~\ref{fig:architecture}).

\noindent\textbf{Modality Fusion}:
A simple way for feature fusion is through channel concatenation~\cite{jiang2018w}. However, this strategy does not really merge the modality-specific information into modality-independent features, so that the contribution of different modalities can not be balanced dynamically.
Such adaptive balancing among modalities is particularly important in pathology segmentation, where specific pathologies can only be spot in particular modalities, i.e. infarct can only be seen in LGE, while edema can only be seen in T2, as seen in Fig.~\ref{fig:challenge}.

Instead, we would like to fuse the feature in an auto-selective fashion. 
To this end, we employ the pixel-wise maximum operator, which has been previously used in~\cite{chartsias2019disentangle} for dual-modal anatomy segmentation.
In the proposed MFU-Net, the fusion is among features generated by the dedicated encoders, producing the fused feature as depicted in pink blocks in Fig.~\ref{fig:architecture}.
Rather than fusing latent features of one layer~\cite{chartsias2019disentangle}, we apply the max-fusion operation to different blocks in the encoders for multi-scale mixture of the multi-modal information.
For instance, for the $k$-th encoding layer, the fusion is performed by $Enc^k(x^i_{LGE},x^i_{T2},x^i_{bSSFP})=\max(Enc^k_{LGE}(x^i_{LGE}),  Enc^k_{T2}(x^i_{T2}), Enc^k_{bSSFP}(x^i_{bSSFP}))$ in a pixel-wise fashion.\footnote{For simplicity we note it as $Enc^k$ in following sections.} It provides the dynamically selective features across modalities. However, the conventional concatenation features do not differentiate features from different modalities.
The fused feature $Enc^k$, together with the linear concatenation of $Enc^k_{LGE}(x^i_{LGE})$,  $Enc^k_{T2}(x^i_{T2}$, and $Enc^k_{bSSFP}(x^i_{bSSFP})$, are then concatenated to the corresponding decoding layer with a skip connection, as in the original U-Net~\cite{ronneberger2015u}. The linear concatenated and nonlinear max-fused representations provide the complementary information for the modal-specific features.

\begin{figure}[t!]
\begin{subfigure}{0.5\linewidth}
  \centering
  \includegraphics[width=0.75\linewidth]{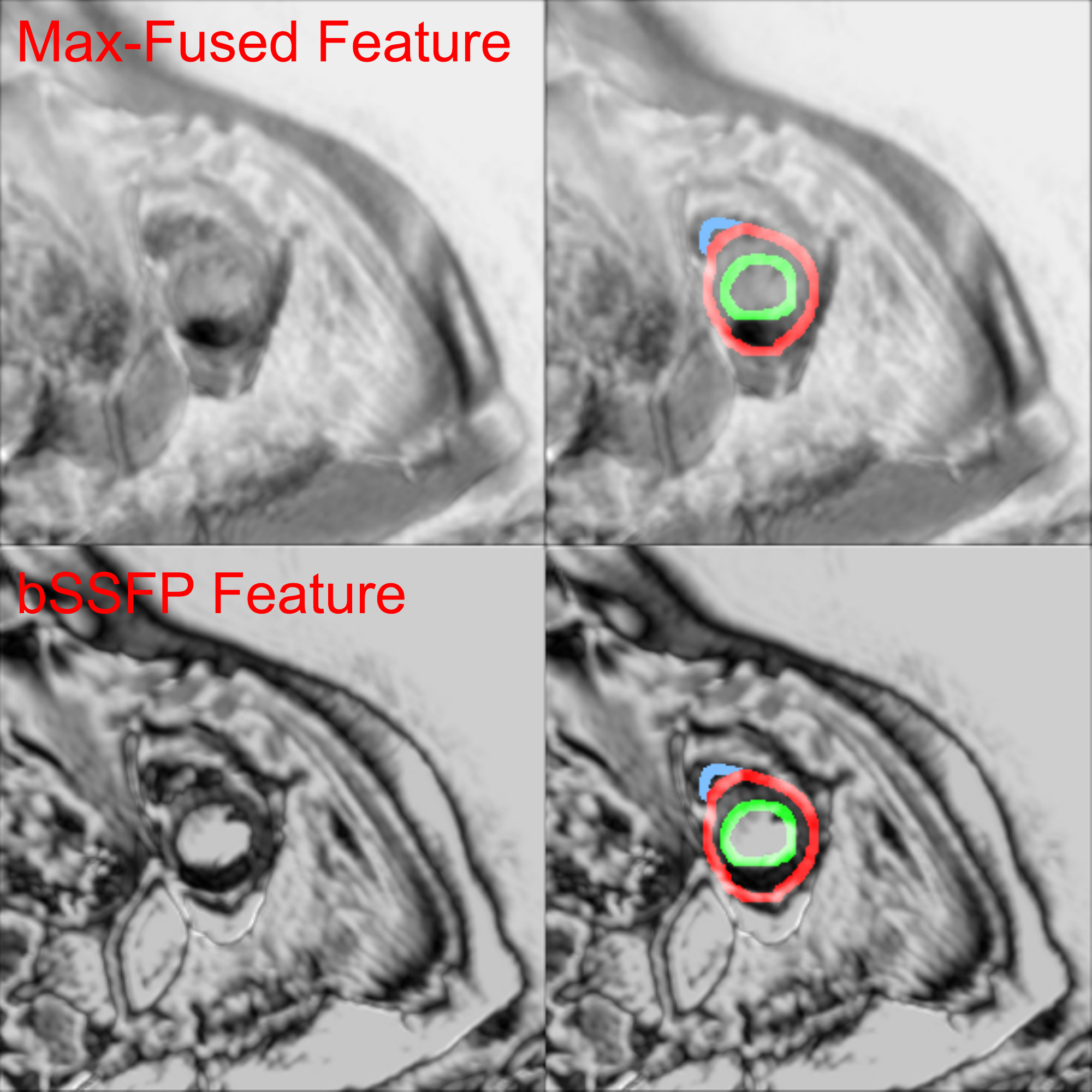}
  \caption{Example 1 with anatomy overlay}
  \label{fig:anato1}
\end{subfigure}
\begin{subfigure}{0.5\linewidth}
  \centering
  \includegraphics[width=0.75\linewidth]{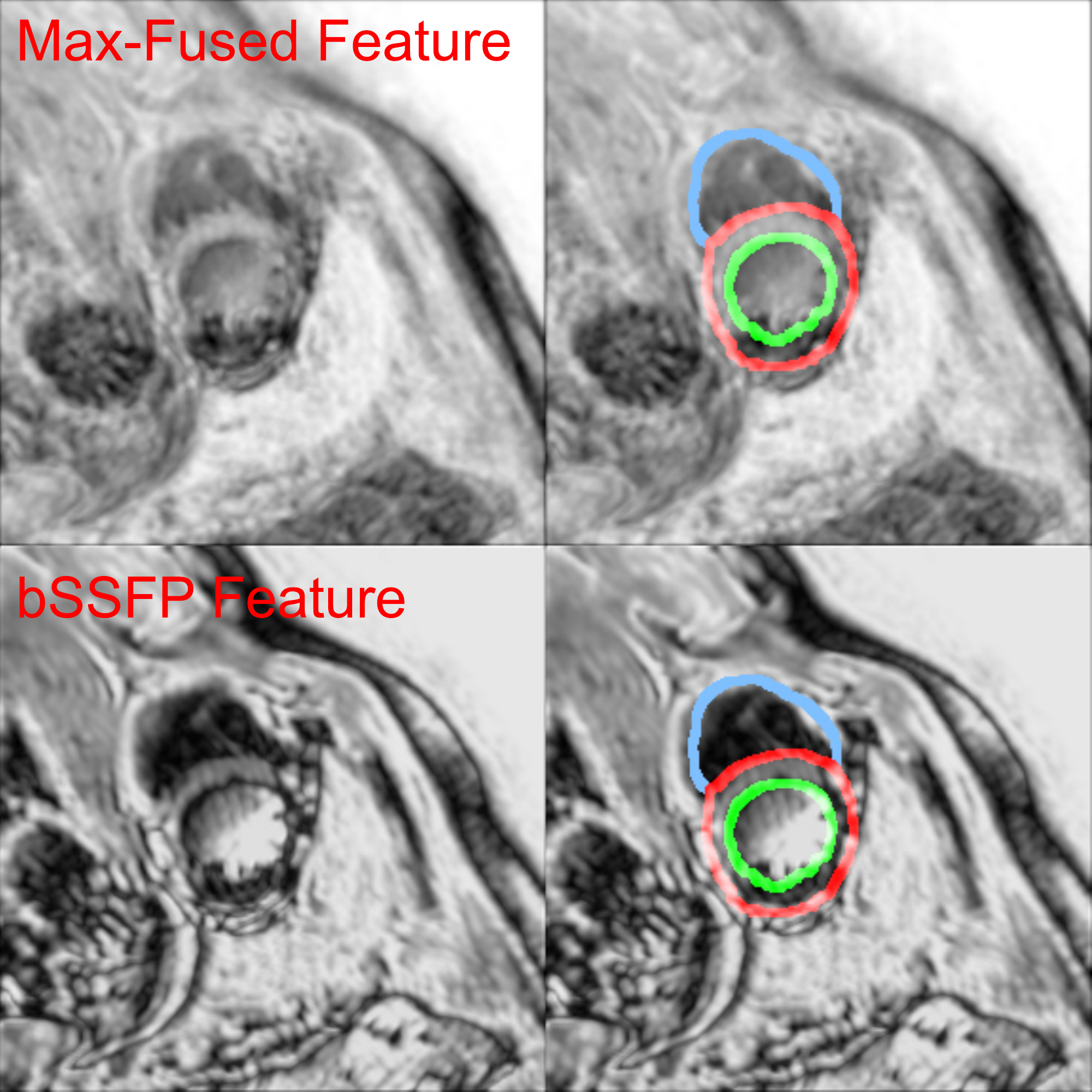}
  \caption{Example 2 with anatomy overlay}
  \label{fig:anato2}
\end{subfigure}
\begin{subfigure}{0.5\linewidth}
  \centering
  \includegraphics[width=0.75\linewidth]{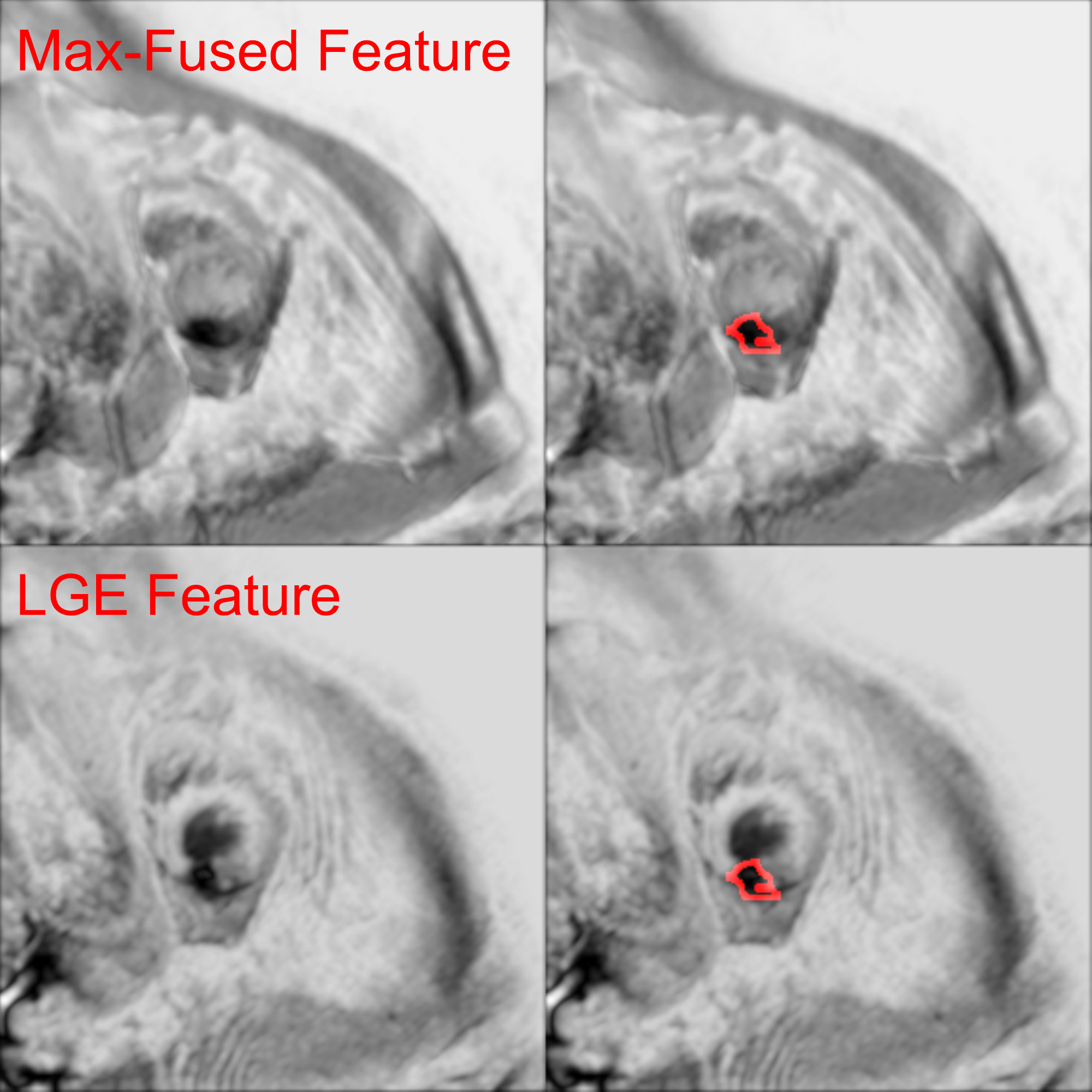}
  \caption{Example 1 with infarct overlay}
  \label{fig:infarct1}
\end{subfigure}
\begin{subfigure}{0.5\linewidth}
  \centering
  \includegraphics[width=0.75\linewidth]{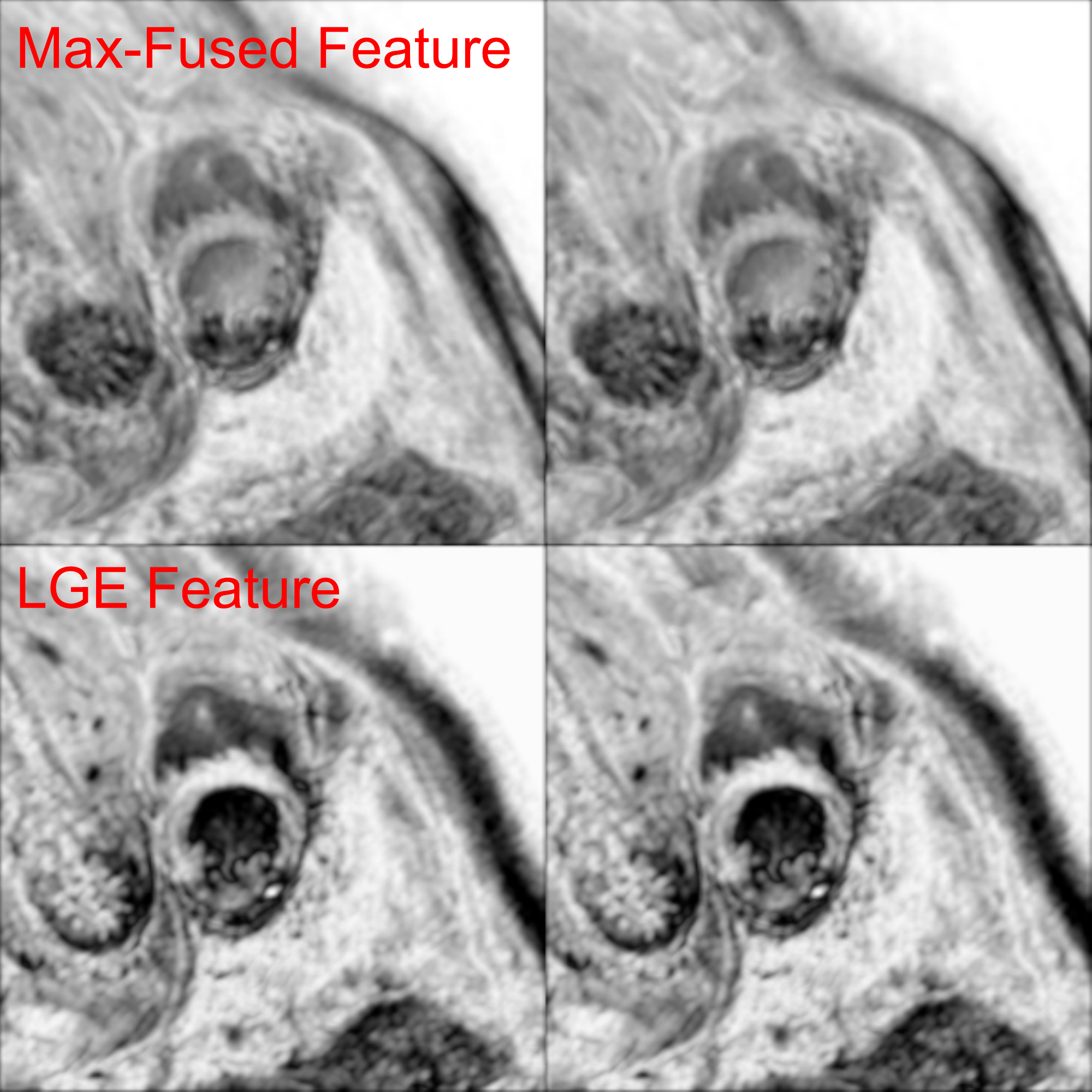}
  \caption{Example 2 with infarct overlay}
  \label{fig:infarct2}
\end{subfigure}
\begin{subfigure}{0.5\linewidth}
  \centering
  \includegraphics[width=0.75\linewidth]{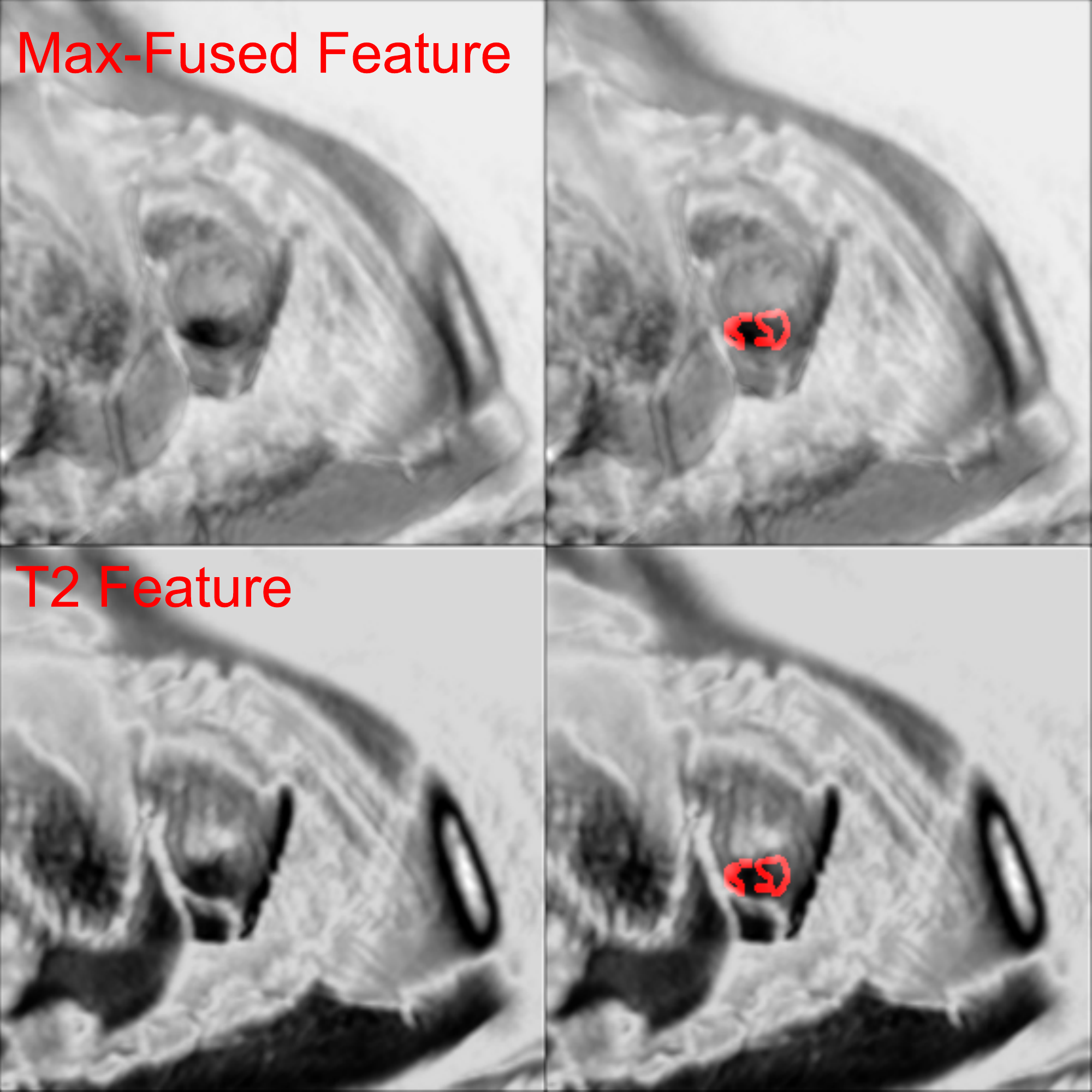}
  \caption{Example 1 with edema overlay}
  \label{fig:edema1}
\end{subfigure}
\begin{subfigure}{0.5\linewidth}
  \centering
  \includegraphics[width=0.75\linewidth]{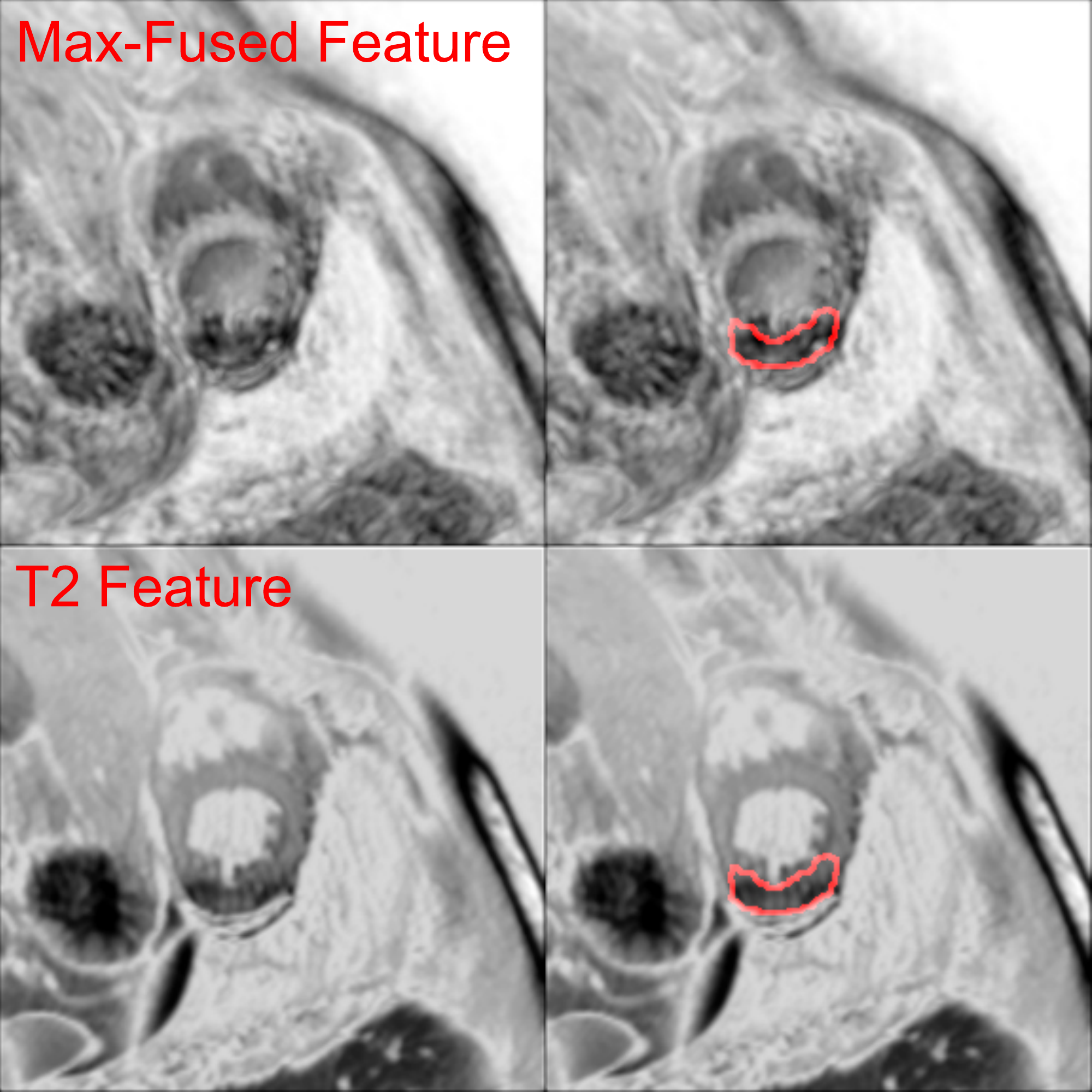}
  \caption{Example 2 with edema overlay}
  \label{fig:edema2}
\end{subfigure}
\caption{Two examples of comparison between the feature maps extracted before and after the max-fusion operation in terms of visibility of: (a) and (b) anatomy; (c) and (d), myocardial infarction; (e) and (f) edema. For each subfigure, the max-fused feature maps are shown at the top and modality-specific feature maps are shown at the bottom. The object boundaries overlapped with the feature maps are on the right. }
\label{fig:maxfuse}
\end{figure}

Two examples of the max-fused features compared to single-modal features are shown in Fig.~\ref{fig:maxfuse}. As discussed above, specific pathologies can only be observed in particular modalities clearly. For example, myocardial infarction can only be observed on features extracted from LGE data as a small dark area (Fig.~\ref{fig:infarct1} and~\ref{fig:infarct2}). Similarly, the boundary of edema can only be depicted on T2 feature maps (Fig.~\ref{fig:edema1} and~\ref{fig:edema2}). In comparison, both pathological regions can be easily detected with relatively clearer boundaries on the max-fused feature maps (Fig.~\ref{fig:infarct1},~\ref{fig:infarct2},~\ref{fig:edema1},~\ref{fig:edema2}). Furthermore, the interested anatomical structures can be seen as easily as in bSSFP features (Fig.~\ref{fig:anato1} and~\ref{fig:anato2}). On the contrary, the boundaries of heart anatomy and edema are blurred in LGE, so is the infarction in T2 data. Boundaries of both infarction and edema are hard to be detected in bSSFP data. This can be seen as a qualitative evidence of an effective mixture of the multi-modality information.


\noindent\textbf{Decoding with Attention}:
The decoder of MFU-Net 
receives as input the bottleneck layer that follows the concatenated multi-modal features of the encoding part. 
A series of convolutional blocks upsample the spatial resolution as in U-Net, and are concatenated with the encoding features (including the max-fused feature and the corresponding encoding features for each modality) computed at the corresponding layers of the encoder with skip connections.

\begin{figure}[t!]
    \centering
    \includegraphics[width=\linewidth]{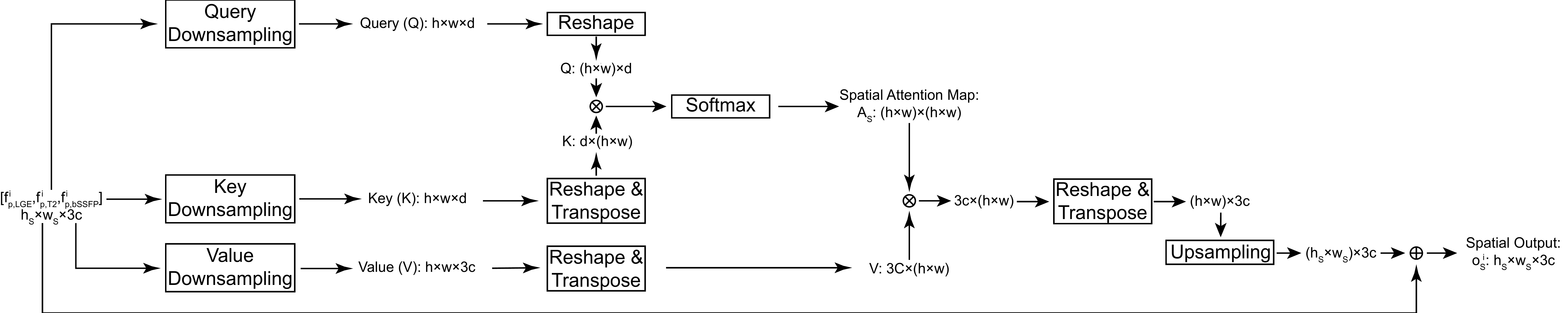}
    \caption{Attention Module at the last decoding layer. $\oplus$ and $\otimes$ represent element-wise summation and multiplication respectively between two matrices.}\label{fig:attention}
\end{figure}

\begin{figure}[t!]
    \centering
    \includegraphics[width=\linewidth]{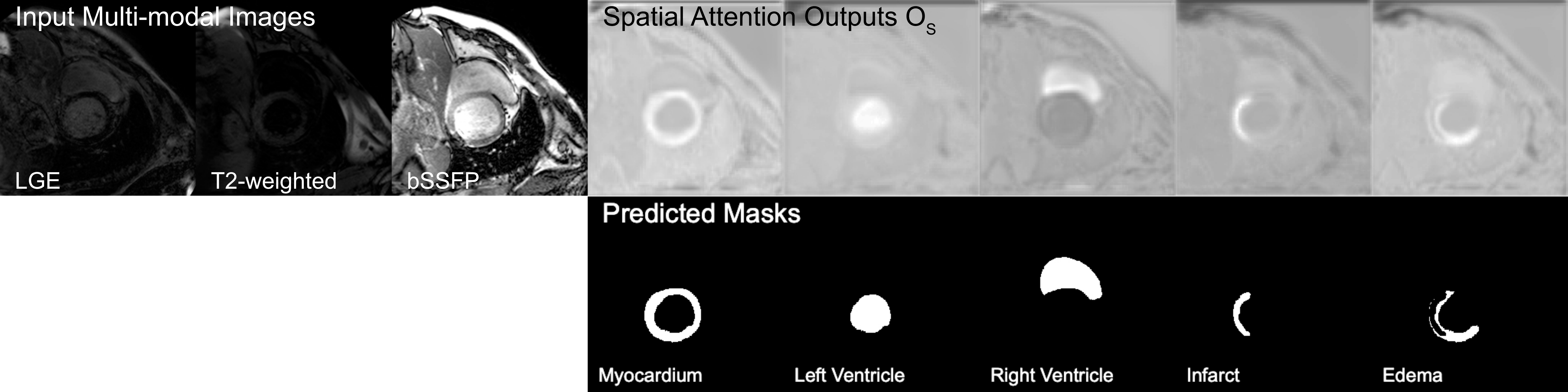}
    \caption{Spatial attention outputs correspond to the predicted masks. }\label{fig:attention-map}
\end{figure}

Since cardiac pathologies often occupy in a small part of the whole image, producing segmentations by treating each pixel equally is challenging and might lead the network to concentrate more on the background but ignore tiny pathological regions. 
In order to overcome this issue, we use a spatial attention mechanism~\cite{fu2019dual} to capture long-range pixel dependencies and assign different weights on different regions.
In this sense, segmentation can be improved by selecting useful information in features extracted around the pathological regions and by discarding unrelated features.
In detail,  the spatial attention module, shown in Fig.~\ref{fig:attention}, is applied at the last layer of the decoding path with the architecture of the spatial and channel attention modules following~\cite{fu2019dual}.
In order to reduce computational complexity introduced when the feature dimensions are large, we first downsample the input feature using stride-2 convolutions before calculating the query, key, and value tensors. The attention module is depicted in Fig.~\ref{fig:attention}.
After calculating the attention map, the dimension will be recovered by deconvolution in the upsampling block.
Fig.~\ref{fig:attention-map} gives examples of spatial attention outputs with corresponding predicted masks. 
Clearly, the corresponding mask region is highlighted in the spatial attention maps, demonstrating the utility of this mechanism in segmentation.

\section{Implementation} \label{sec:implementation}
In this section, the implementation details of the proposed MFU-Net will be specified.
Firstly, we will introduce the dynamic resampling training strategy, then the alternative cross-validation to make full use of the training data and avoid overfitting issue will be specified.
\subsection{Dynamic Resampling Training Strategy}
\begin{figure}[t!]
    \centering
    \includegraphics[width=\linewidth]{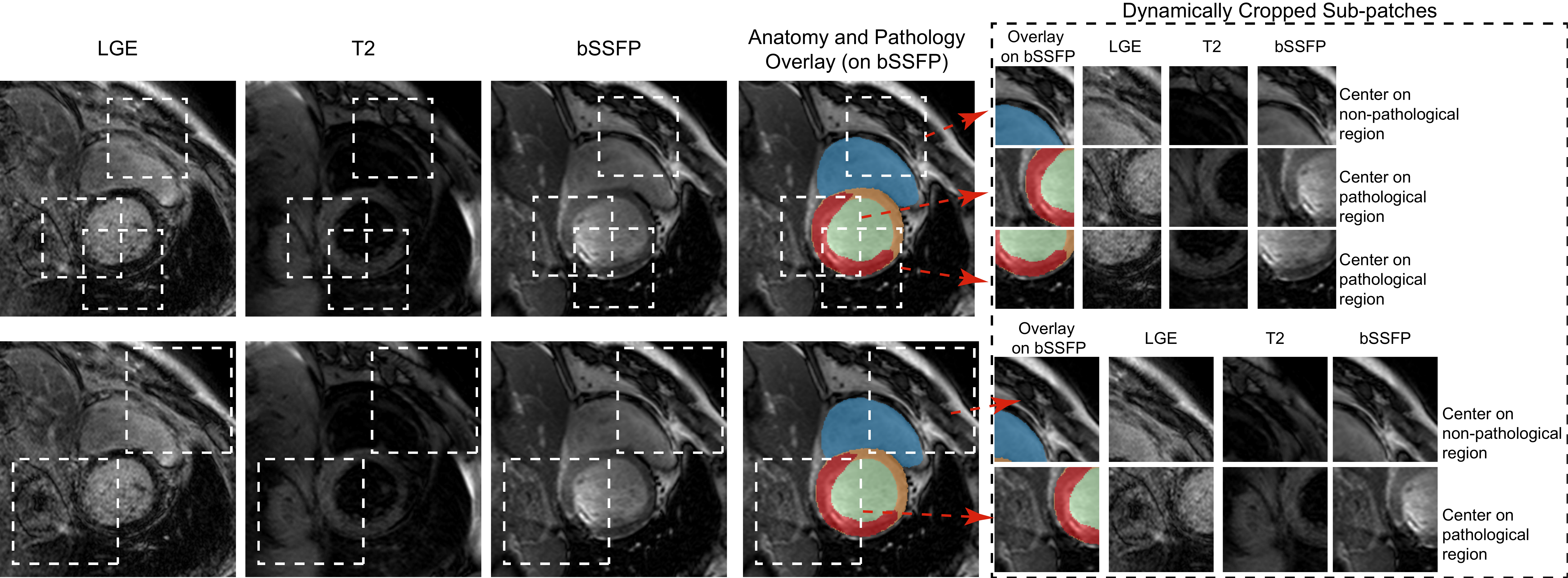}
    \caption{Dynamic resampled image patches with varying spatial and batch sizes. The resampling sizes for images in the first and second row is 96 and 128 respectively. Smaller resampling size will bring greater batch size.}\label{fig:dynamic-sampling}
\end{figure}

The proposed MFU-Net is deployed on a GTX Titan X GPU with 12GB standard memory. In the training process, the available memory allows $288\times 288$ image size with a batch size equals to 4. 
In order to increase the model's focus on pathological regions, we also train with patches of different sizes that are dynamically resampled.
For the batch obtained at the $t$-th iteration, we first decide the patch size $d_t$ by $d_t = 96+16i$, $i \in \{ 1, \cdots, 12 \}$ where $i$ is randomly picked. Then, with an arbitrary probability $\rho_{c}$, an extracted patch is centred on the pathology of interest. The dynamic batch size $N_{t}$ is decided by $N_{t} = \lfloor d_{t-1}^2 N_{t-1} / d_{t}^{2} \rfloor$. For example, in the first iteration, we initialize the image size $d_{0} = 288$, thus when extracting $96\times 96$ image patches, the batch size can be as big as 36. This not only increases the batch size but also allows to manual balance the data distribution. In this work, we set $ \rho_{c}=0.89 $ as the interested anatomy only takes up $~11\%$ pixels of the whole image. As such, pathological regions are more probably to be seen in the cropped patches. Fig.~\ref{fig:dynamic-sampling}
demonstrates the details of this sampling process with two different patch sizes.

\subsection{Training with Alternative Cross Validation}~\label{sec:alternative-cross-validation}
To make full use of the training data and avoid possible overfitting issues, we employ an alternative cross-validation strategy as part of training to predict the MyoPS 2020 challenge testing data.
Specifically, the whole training set is split into five parts.
Accordingly, the training process will be specified in five phases.
In each phase, four out of the five splits are selected as the training set, while the remaining one is used as the validation set to prevent overfitting by defining the early-stopping criteria.
If one phase of training is terminated, the network optimization continues on another split.
The number of epochs for each training phase are 50, 40, 30, 20, and 15, while the initial learning rate are 0.0001, 0.00009, 0.00008, 0.00006, 0.00005 respectively and decayed exponentially.
When all the five training phases are completed, we add a final fine-tuning phase that involves all the training data but is trained only 10 epochs with the small learning rate at 0.00004 and decayed exponentially as well. This will avoid the model to forget early trained examples.

\section{Experiments}
We evaluate the proposed MFU-Net on pathology segmentation using the Dice score. 
Experimental setup, datasets, benchmarks, and training details will be detailed in the following part.

\noindent\textbf{Data:}
We evaluate our proposed MFU-Net on the multi-sequence CMR (MyoPS) dataset~\cite{zhuang2016multivariate,zhuang2018multivariate} that contains in total 25 volumes and 102 slices in the training set.
For each slice, three modalities including LGE, T2, and bSSFP are provided. They are preprocessed with the Multi-variate Mixture Model~\cite{zhuang2016multivariate,zhuang2018multivariate}, such images from the three modalities are aligned and resampled to same spatial resolution.
For all the images, three anatomy masks (myocardium, left ventricle, and right ventricle) and two pathology masks (myocardial infarct and edema) are given.
The testing set contains 20 volumes and 72 slices without ground-truth masks available.
Both training and testing data are cropped to $288\times288$ to keep the region to be segmented in the sight.

\noindent\textbf{Training details:}~\label{sec:training-detail}
The proposed MFU-Net is optimized with fully supervised losses. 
The segmentation of both anatomy and pathology is trained with tversky~\cite{salehi2017tversky} and focal~\cite{lin2017focal} losses in a supervised fashion.
The tversky loss is defined as $\ell_{T,j}=(\hat{y}^i_{j} \odot y^i_{j}) /[ \hat{y}^i_{j} + y^i_{j} + (1-\beta)\cdot(\hat{y}^i_{j}-\hat{y}^i_{j} \odot y^i_{j})+\beta\cdot(y^i_{j}-\hat{y}^i_{j} \odot y^i_{j})]$ and the focal loss is $\ell_{F,j} = \sum_{H,W}[-y^i_{j}(1-\hat{y}^i_{j})^\gamma log(\hat{y}^i_{j})]$, where
$\odot$ represents the element-wise multiplication and $j$ corresponds to the involved anatomy or pathology labels.
We set penalties for anatomy, infarct, and edema equal to $\lambda_{anatomy}=1$, $\lambda_{infarct}=3$, and $\lambda_{ana}=5$ respectively, for each of the tversky and focal losses.
Moreover, in order to achieve more stable training and quicker convergence, we initialise MFU-Net with weights from the MMSDNet~\cite{chartsias2019disentangle} encoder (that also follows a U-Net architecture with dedicated encoders for each modality) when trained only with the unsupervised reconstruction loss.

\noindent\textbf{Benchmarks:}
We evaluate the pathology segmentation performance of MFU-Net using several variants of our model.
More specifically, we evaluate the effect of different design choices including the maximum fusion operator, the spatial attention module and the dynamic resampling strategy. In total we construct eight ablated models, all of which concatenate features at each encoding layer.


\begin{table}[t!]
\caption{Anatomy and pathology segmentation dice scores (\%) of MFU-Net and relevant variants with \textit{Residual} backbone. Myo., LV, and RV represent the myocardium, left ventricle, and right ventricle respectively. \textit{max}, \textit{attention}, and \textit{resample} represent the presence of the max-fusion operator, the spatial attention module, and the dynamic resampling strategy respectively. Pathology score is calculated by averaging both the infarct and edema segmentation performance. }\label{tab:residual}
\centering
\begin{tabular}{ccc|ccc|cc|c}
\hline
\textit{max} & \textit{attention} & \textit{resample} 
& Myo. & LV & RV & Infarct & Edema & Avg. Pathology   \\\hline
\checkmark & \checkmark & \checkmark
             &$84.3_{7.9}$
             &$\mathbf{87.5_{7.1}}$ 
             &$78.5_{14.2}$
             &$\mathbf{53.0_{20.5}}$
             &$28.7_{13.9}$
             &$\mathbf{44.9_{13.9}}$\\
-- & \checkmark & \checkmark     
             &$\mathbf{85.2_{8.1}}$
             &$86.8_{10.6}$
             &$\mathbf{78.7_{14.0}}$
             &$52.1_{20.4}$
             &$\mathbf{29.4_{12.4}}$
             &$42.9_{14.0}$\\
\checkmark & -- & \checkmark     
             &$84.2_{6.9}$
             &$86.9_{7.5}$
             &$76.7_{13.7}$
             &$46.1_{21.1}$
             &$28.1_{14.3}$
             &$41.0_{14.1}$\\
\checkmark & \checkmark & --
             &$84.5_{5.3}$
             &$87.1_{6.4}$
             &$74.9_{18.7}$
             &$49.4_{20.4}$
             &$\mathbf{29.4_{17.9}}$
             &$42.8_{15.5}$ \\
-- & -- & \checkmark             
             &$81.1_{7.8}$
             &$84.2_{8.0}$
             &$67.2_{17.2}$
             &$50.2_{17.8}$
             &$19.3_{13.0}$&$37.5_{16.8}$\\
-- & \checkmark & --
             &$\mathbf{85.2_{4.1}}$
             &$86.1_{9.4}$ 
             &$75.7_{18.6}$ 
             &$52.6_{19.4}$ 
             &$28.7_{17.1}$ 
             &$43.6_{15.4}$ \\
\checkmark & -- & --             
             &$82.3_{7.9}$
             &$82.3_{8.9}$ 
             &$68.2_{16.5}$ 
             &$48.0_{25.4}$ 
             &$22.8_{15.9}$ 
             &$36.1_{18.5}$ \\
-- & -- & --
             &$81.6_{6.5}$
             &$84.1_{8.0}$ 
             &$67.5_{15.5}$ 
             &$42.8_{21.7}$ 
             &$20.6_{16.6}$ 
             &$34.8_{17.7}$ \\\hline
\end{tabular}
\end{table}

\subsection{Results and Discussion} \label{sec:results}
We report segmentation results of MFU-Net and the ablated models in Table~\ref{tab:residual} with anatomy (myocardium, left and right ventricles) and pathology (myocardial infarct and edema) segmentation dice scores.\footnote{Since we do not have the ground truth of the testing data, the performance reported in Table~\ref{tab:residual} and Table~\ref{tab:backbone-comparison} are obtained by five-fold cross validation across the training set. Relevant splits are following the description in Sec.~\ref{sec:alternative-cross-validation}. In addition, we also report the averaged pathology Dice scores of the both pathologies to assess the overall pathology segmentation performance.}
The backbone architecture used the residual connections in encoding and decoding layers~\cite{he2016deep} noted as \textit{Residual}.

As can be seen in Table~\ref{tab:residual}, the proposed maximum fusion operator and dynamic resampling achieve the best infarct segmentation, while edema segmentation performs similarly to the model without the max fusion. On average the model with all \textit{attention}, \textit{max}, and \textit{resample} options achieves the best pathology segmentation with Dice equal to 44.9\%.\footnote{Although the anatomy segmentation performance decreases, we still think \textit{SideConv} and \textit{Dilation} are better choices since we are more caring about the pathology prediction in this research.}
Moreover, it can be observed that the spatial attention module improves segmentation for both infarct and edema.

In addition, the anatomy segmentation does not benefit from the proposed compositions, particularly in ventricles. 
The reason is two-folded. On one hand, the MyoPS 2020 challenge concentrates mainly on the pathology segmentation. As such, during training, we put more penalties on the pathology supervision (Sec.~\ref{sec:training-detail}). It results in less focus on anatomy learning.
On the other hand, because both infarct and edema is in the myocardium region, the pathology training gradient will offer an additional guide to train myocardium segmentation. 
On the contrary, ventricle predictions are not enjoying such an advantage.

\subsection{Prediction for the Challenge Testing Dataset}

\begin{table}[t!]
\caption{Anatomy and pathology segmentation comparison between \textit{Residual}, \textit{Dilation}, and \textit{Sideconv} backbones when \textit{max}, \textit{attention}, and \textit{resample} are all present.}\label{tab:backbone-comparison}
\centering
\begin{tabular}{c|ccc|cc|c}
\hline
         & Myo. & LV & RV & Infarct & Edema & Avg. Pathology \\ \hline
\textit{Residual} 
             &$\mathbf{84.3_{7.9}}$
             &$\mathbf{87.5_{7.1}}$ 
             &$\mathbf{78.5_{14.2}}$
             &$53.0_{20.5}$
             &$28.7_{13.9}$
             &$\mathbf{44.9_{13.9}}$\\ \hline
\textit{Dilation} 
             &$80.5_{4.3}$
             &$85.3_{6.3}$ 
             &$44.3_{33.8}$
             &$\mathbf{55.1_{18.7}}$
             &$23.1_{13.9}$
             &$43.7_{14.0}$\\ \hline
\textit{Sideconv} 
             &$76.3_{10.3}$
             &$65.0_{18.6}$ 
             &$40.5_{39.4}$
             &$52.1_{21.1}$
             &$\mathbf{29.7_{11.8}}$
             &$45.0_{16.0}$\\ \hline
\end{tabular}
\end{table}

Table~\ref{tab:backbone-comparison} specifies the comparison with other two backbone CNN options, namely, the dilated convolutions in the bottleneck layer~\cite{vesal20192d}, and the side-convolution by adding $3\times3$, $3\times1$, and $1\times3$ convolutions in each of the convolution operations~\cite{ding2019acnet}.
They are denoted as \textit{Dilation} and \textit{SideConv} respectively.
It can be seen clearly that the models using dilated convolutions and side-convolutions improve on the segmentation of infarct and edema respectively, compared to our initial model using residual connections. 
We therefore use the \textit{Dilation} and \textit{Sideconv} MFU-Nets for inference of the MyoPS 2020 testing dataset. The segmentation results are presented in Table~\ref{tab:test} and contain the Dice scores of infarct and the union of both infarct and edema. It can be seen that the \textit{dilation} backbone with \textit{max}, \textit{attention}, and \textit{resample} achieves better results with 58.4\% dice for infarct, and 61.4\% for both the infarct and the edema together.

\begin{table}[t!]
\caption{Pathology segmentation dice scores on the MyoPS 2020 testing data}\label{tab:test}
\centering
\begin{tabular}{cc|cc}
\hline
\multicolumn{2}{c|}{SideConv} & \multicolumn{2}{c}{Dilation} \\ \hline
Infarct       & Infarct+Edema    & Infarct       & Infarct+Edema\\ \hline
$57.0_{28.7}$ & $60.3_{18.1}$    & $58.4_{26.3}$ & $61.4_{17.8}$\\ \hline
\end{tabular}
\end{table}

\begin{figure}[ht]
\begin{subfigure}{0.5\linewidth}
  \centering
  \includegraphics[width=\linewidth]{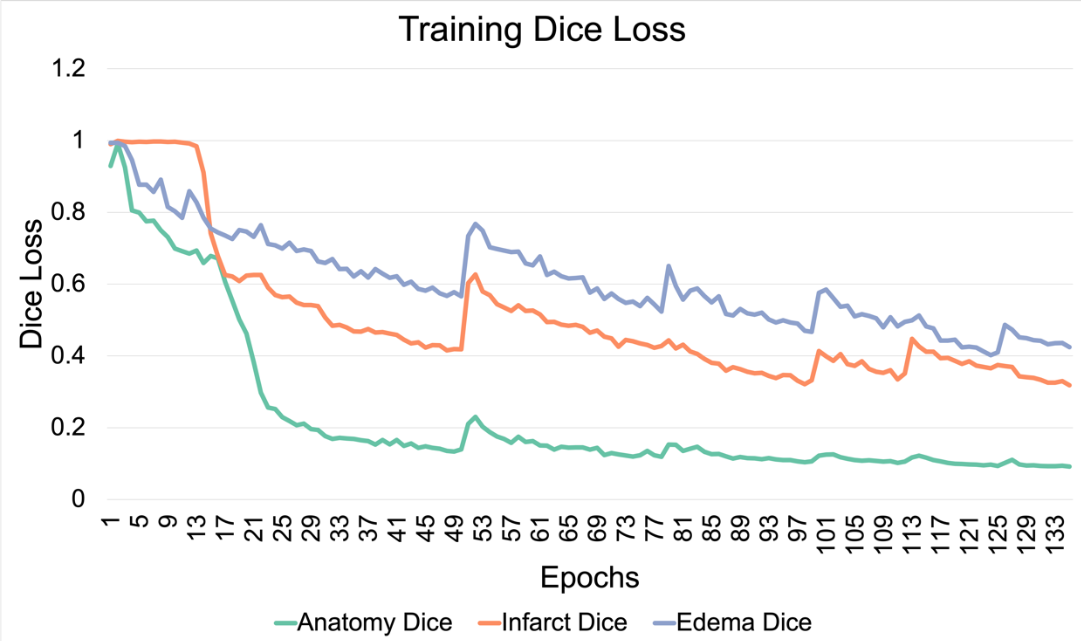}
  \caption{Training dice losses}
  \label{fig:trainingdice}
\end{subfigure}
\begin{subfigure}{0.5\linewidth}
  \centering
  \includegraphics[width=\linewidth]{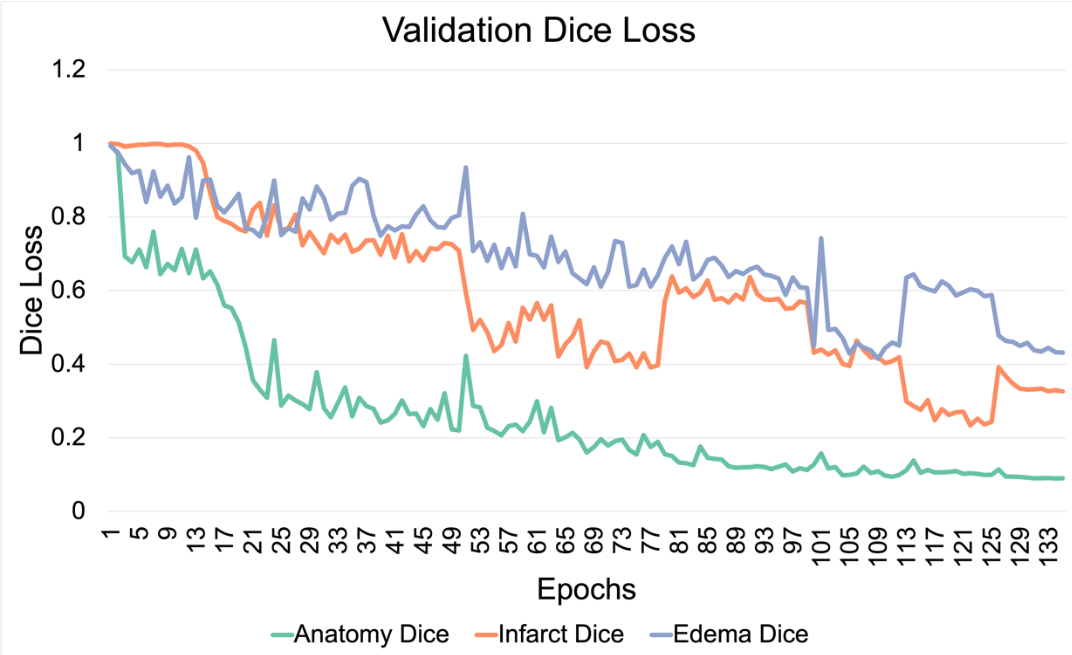}
  \caption{Validation dice losses}
  \label{fig:validatingdice}
\end{subfigure}
\caption{Training and validation dice losses with the alternative cross-validation for the testing dataset. Curves in green, orange, and blue represent the anatomy, infarct, and edema dice losses.}
\label{fig:validationdice}
\end{figure}

The prediction models are trained with the alternative cross validation described in Sec.~\ref{sec:alternative-cross-validation}.
Fig.~\ref{fig:trainingdice} and Fig.~\ref{fig:validatingdice} illustrate the training and validation dice losses respectively during model optimization.
Particularly, in Fig.~\ref{fig:trainingdice}, each loss jump corresponds to the point where the cross validation split switches and the training phase changes. Furthermore, all losses gradually decrease in each training phase, and finally converge at the final few steps.

\section{Conclusions}
In this paper, we proposed the Multi-Fusion U-Net, a novel architecture to segment infarct and edema from multi-modal images including LGE, T2-weighted, and bSSFP sequences.
Our model uses dedicated encoders for each modality, and combines multi-modal information with feature fusion performed with the pixel-wise maximum operator at each encoding layer.
These max-fused features together with the concatenated modality-specific features of each encoding layer, are propagated to corresponding decoding layers of the same spatial resolution using skip connections.
Additionally, a spatial attention module in the final decoding layer, as well as a novel dynamic resampling training strategy, are engaged to guide the network to focus on small pathology regions.
Extensive experiments on the MyoPS 2020 challenge dataset demonstrated the effectiveness of the MFU-Net in improving cardiac pathology segmentation performance.

~\\
\noindent\textbf{Acknowledgement:} This work was supported by US National Institutes of Health (1R01HL136578-01). This work used resources provided by the Edinburgh Compute and Data Facility (http://www.ecdf.ed.ac.uk/).
S.A. Tsaftaris acknowledges the Royal Academy of Engineering and the Research Chairs and Senior Research Fellowships scheme. 

\clearpage
\bibliographystyle{splncs04}
\bibliography{miccai}

\end{document}